\documentclass[12pt,aps,bm,epsfig,amssymb]{revtex4}
\usepackage{epsfig,amsmath,amsfonts,float}
\newcommand{\R}[0]{{\mathbb{R}}}

\def\C{{\mathbb{C}}}

\def\R{{\mathbb{R}}}

\def\N{{\mathbb{N}}}

\newcommand{\pha}{ 2 \pi i }
\newcommand{\bef}{\begin{figure}}
\newcommand{\eef}{\end{figure}}

\newcommand{\leb}{\left(}
\newcommand{\rib}{\right)}
\newcommand{\bei}{\begin{itemize}}
\newcommand{\eei}{\end{itemize}}
\newcommand{\bea}{\begin{eqnarray}}

\newcommand{\eea}{\end{eqnarray}}
\newcommand{\bequ}{\begin{equation}}
\newcommand{\eequ}{\end{equation}}

\DeclareRobustCommand\openone{\leavevmode\hbox{\small1\normalsize\kern-.33em1}}

\begin{document} 

\title{On Quantum A/D and D/A Conversion}

\author{Frank Schm\"user$^1$ and Dominik Janzing$^{1, \, 2}$}
\affiliation{$^1$ Institut f\"ur Algorithmen und Kognitive Systeme, 
Universit\"at Karlsruhe,
Am Fasanengarten 5, 76131 Karlsruhe, Germany. \\
$^2$ Institut f\"ur Quantenoptik und Quanteninformation, \"Osterreichische Akademie der Wissenschaften,  Technikerstr.~25, 6020 Innsbruck, Austria.}
\date{May 2, 2005}

\begin{abstract}
An algorithm is proposed which transfers the quantum information of a wave function
(analogue signal) into a register of qubits (digital signal) such that $n$ qubits describe
the amplitudes and phases of $2^n$ points of a sufficiently smooth wave function. 
We assume that the continuous degree of freedom couples to one or more qubits of a
quantum register via a Jaynes Cummings Hamiltonian and that we have universal
quantum computation capabilities on the register as well as the possibility to
perform bang-bang control on the qubits. The transfer of 
information is mainly based on the application of the quantum phase-estimation algorithm in both directions. 
Here, the running time increases exponentially with the number of qubits. We pose it as an open question
which interactions would allow polynomial running time. One example would be interactions which 
enable squeezing operations. 
\end{abstract}

% The well--known phase 
%estimation algorithm plays an important role in our scheme. Furthermore, it is shown that 
%the analogue to digital converter can be realised in the context of the ion 
%trap quantum computer, since the necessary operators can be generated by an interplay between
%the Jaynes Cummings Hamiltonian and one qubit control operations. The A/D converter
%seems to be a useful device for hybrid quantum computation that uses both discrete and 
%continuous quantum variables as resources. 
%\end{abstract}
%
\maketitle
\section{Introduction}
Traditionally quantum computing and quantum cryptography 
have been formulated in a digital setting, i.e.~with qubits 
\cite{NC}.
However, also models with continuous
variables have been proposed \cite{ContinuousLloyd}.
Protocols for continuous variable cryptography have  been investigated in detail (e.g. \cite{Silberhorn}).
In \cite{loyll} several operations on hypothetical 
continuous variable quantum computers have been proposed which 
can generate arbitrary unitaries. The author argues that continuous
models possess various advantages compared to the standard model quantum
computer. Therefore an interface between continuous and discrete
registers is desirable, since with this device one could combine 
the advantages of both approaches. 
There are also other reasons why 
the bridge between continuous and discrete degrees of freedom
is an interesting issue of research: The possibility to  transfer the  wave function of a massive particle
or the state of a light mode to a quantum register would allow to use algorithmic measurement schemes 
like those proposed in  \cite{deck}  for POVM measurements on the continuous degree of freedom. 
Similarly, the ability to transfer quantum information from digital to analogue
would allow to use state preparation algorithms in 
quantum computers \cite{Soklakov} for  algorithmic state generation
in the analogue system. Furthermore, the implementation of POVM measurements with an uncountable number of outcomes
on a finite dimensional system is only possible if one couples it to a continuous degree of freedom \cite{Ariano}. 

An interesting system where the state of a light field is transfered to the state of many two-level systems
and vice versa is the micromaser (see \cite{wellens} and references therein). 
The two-level atoms cross a cavity one after another such that
at most one atom is present in the cavity at any time. While it is passing the cavity, each atom
is interacting with the cavity field mode via a Jaynes Cummings Hamiltonian.  One can prove \cite{wellens}
that every state of the field mode can asymptotically 
be prepared as a limit if an infinite number of atoms, initialized
to an appropriate state, passes the cavity. It has been shown that 
for many interesting examples  small numbers of atoms  
are sufficient to prepare the desired state with high fidelity. Since the final state of the field
in the asymptotic scheme does not depend on its initial state, the latter has been 
completely transferred to the outgoing atoms. Therefore the system realizes {\it asymptotically} the 
transfer of quantum information in both directions. However, the fact that
these statements refers to asymptotic behaviour indicates already that 
the state is typically not encoded on a minimal number of atoms. 

Another system where  quantum state transfer between a multi-photon state and the states of atoms 
and vice versa has already been experimentally implemented is described in \cite{Polzik}. 
In this ``quantum memory for light'' the eigenvalues of the total spin operator $J_z$ of the atoms 
define the basis states of the 
atomic memory. 
However, this scheme encodes an $n$-photon states
in a collective polarization of an atom ensemble where the number of atoms is also of the order $n$, i.e.,
the number of qubits is in the order of the dimension of the encoded space. 
In this article we propose a quantum analogue-to-digital converter where the number of qubits
needed grows only logarithmically in the dimension of the encoded space for the cost of an 
exponential running time of the algorithm. We shall discuss later whether 
this shortcoming can be removed. The question of the cost 
of accurate A/D conversion 
is directly connected with
the question of the computational power  of analogue computers, which is
already an interesting problem in classical computer science \cite{Vergis}. 
Whether or not a continuous degree of freedom 
could be used to store a ``reasonable number'' of qubits depends on 
the ability to access a subspace of exponentially large dimension on a ``reasonable'' time scale.

Here the continuous degree of freedom is represented mathematically by
the Hilbert space $L^2(\R)$, the set of square integrable functions on the real line.  It is isomorphic
to $l^2(\N_0)$, the space of square summable sequences over $\N_0$ 
by choosing the eigenfunctions
of a harmonic oscillator as complete orthogonal system. This shows 
that continuity or discreteness is here not a property of the Hilbert
spaces but rather of the considered observables.  
Our A/D and D/A converters refer explicitly to a discretization with respect  
to a a variable with continuous spectrum, e.g. the position variable
of a  Schr\"{o}dinger particle, but also applies 
to the formally equivalent  variables of a light mode.
Using the above isomorphism $l^2(\N_0)\equiv L^2(\R)$, 
it would be straightforward to transfer the information such that
the state with $j$ oscillation quanta is mapped onto the $j$th binary word in the discrete register.
However, here we would like to represent the values of the wave functions
at $2^n$ points directly by the coefficients of the binary words of the discrete register.
For doing so, we restrict us to Schr\"odinger 
wave functions that are contained in the interval $[0, \, L]$ (except e.~g.~exponential 
tails). The $n$-qubit register is represented by $(\C^2)^n$ with basis states
$|j\rangle$ with $j=0,\, 1, \, \dots 2^n-1$. 
Then we demand that every sufficiently smooth 
wave function $x \mapsto \psi (x)$ is converted to the quantum register state
\bequ
2^{-n/2} \; \sum_{j=0}^{2^n-1} \psi \Big( \frac{j \; L}{2^n} \, \Big) \; |j\rangle 
\label{consche}
\eequ
in an approximative sense. We will show below that ''sufficiently smooth'' means that the
$L^2$ norm of the derivative of the wave function is not too large. 
The conversion operations that we will use are unitary transformations on
\[
(\C^2)^{\otimes n} \otimes L^2(\R) \quad .
\]

Now we describe the model in which conversion 
from analogue to digital is possible. This provides us with
the available resources for the conversion algorithm.
%The system is a chain of $n$ ions in an r.f. Paul trap that is
%one of the most promising candidates for a quantum computer \cite{CZ,CBla}. 
%A qubit consists of ground and excited state of an ion. Each ion is
%accompanied by its own small laser that induces transitions 
%between its two energy levels. The analogue
%quantum state is the wave function of the phonons that are 
%associated with the centre of mass mode of the ion chain. 

We assume that the interaction is 
 described by the Jaynes 
Cummings Hamiltonian \cite{CBla} 
\bequ
H = c' \, \sum_{j=0}^{n-1} \big( \, \sigma^{(j)}_-  \otimes a^\dagger +
\sigma^{(j)}_+ \otimes a \, \big) \quad ,
\label{jcham}
\eequ
where $c' > 0 $ is the interaction strength and we have used the conventions 

\bequ
a := \frac{1}{\sqrt{2}} \, (\, \hat{x} + i \; \hat{p} \,)\; , \quad 
a^\dagger := \frac{1}{\sqrt{2}} \, (\, \hat{x} - i \; \hat{p} \,) \; , \quad 
\sigma^{(j)}_\pm := \frac{1}{2} \, (\, \sigma_x^{(j)} \pm i \, 
\sigma_y^{(j)} \,) \,,
\eequ
where $\sigma_\alpha^{(j)}$ denotes the Pauli matrix $\sigma_\alpha$ acting on qubit $j$ and
$\hat{x}$ and $\hat{p}$ are the position and momentum operators, respectively, defined by
\[
(\hat{x} \psi)(x) := x\psi(x)  \,\,\,\ , \; \,\,\,(\hat{p}\psi)(x):=-i \, \, \frac{d}{dx} \psi(x) \quad .
\]
We can rewrite the Hamiltonian in  eq.~(\ref{jcham}) as 
\bequ
{H} = c \, \sum_{j=0}^{n-1} \big( \, \sigma^{(j)}_x   \otimes \hat{x} 
- \sigma^{(j)}_y \otimes  \hat{p} \big) \quad .
\label{moham}
\eequ
Note that we choose the oscillator parameters such that $ m \, \omega = 1$ and set $ \hbar = 
1$ throughout the paper. The Hamiltonian (\ref{jcham}) appears often in physical systems when
the continuous degree of freedom is an harmonic oscillator, e.g., an oscillation mode
of ions in a trap \cite{CBla,CZ}. 

To achieve A/D conversion 
it is not sufficient to use just Hamiltonian 
evolution with Hamiltonian (\ref{moham}), we will also need 
various other unitary operators. For example, below we want to use 
the terms $\sigma^{(j)}_x \otimes \hat{x} $ and $\sigma_y^{(j)} \otimes \hat{p} $ 
of eq.~(\ref{moham}) separately. Fortunately, there exists already a well--developed 
technique which allows to simulate various effective Hamiltonians \cite{ernst,Zanardi}. 
Propagating the system only for short time intervals with the Hamiltonian 
(\ref{moham}) and interrupting this by one qubit unitaries, we can cancel or 
modify terms of the Hamiltonian. 
We use the fast control limit (also called bang--bang control), 
i.~e.~Hamiltonian evolution is neglected during one qubit 
operations are applied.
In section \ref{secimp} we will explicitly outline the one qubit 
operations and pulse sequences that entail the desired modifications of the Hamiltonian.
Finally, as our last resource we assume that on the quantum register, 
we have  the ability of universal power of quantum computation. 
Even though we use a specific interaction between continuous and discrete 
register as a resource
for the conversion algorithm there are several  generalizations which will be obvious after having discussed our
method. First, the particle wave needs not necessarily interact with all qubits simultaneously and 
with the same strength, one could also have different coefficients.
 We will furthermore see that the 
only requirement is that one of the interactions
$\sigma_z \otimes \hat{x}$ and  $\sigma_z \otimes \hat{p}$ can be simulated, because
the other can be obtained by implementing a Fourier transform to the continuous system.

We now describe the organization of this article. In section \ref{secconv}
we explain the algorithm that achieves the conversion of quantum information.
Each step of the algorithm is given with its corresponding operator that acts
on the tensor space of qubit register and wave function. In section \ref{secimp}
we describe the procedures for simulation of Hamiltonians which generate the 
required effective Hamiltonians from the given one. 
In section \ref{secDA} we describe briefly that the time reversed implementation
can in principle be used for a digital to analogue converter. 
 We summarize and 
discuss our results in section \ref{secconc}. The appendix gives a proof of 
eq.~(\ref{showbo}).
\section{The A/D conversion algorithm} \label{secconv}
We first sketch the general idea of the functioning of the A/D converter.
It uses a variant of the standard phase estimation algorithm \cite{ClevePhase,NC} 
in order to bring the wave function amplitudes $\psi(x)$ into the appropriate
place of the qubit register (cf.~the scheme of eq.~(\ref{consche})). 
The essential principle is that the interaction $\sigma_z \otimes \hat{x}$
implements a controlled-$\exp(-i \hat{x} T)$ operation which allows to use
the qubit register als
``measurement apparatus'' for  $\hat{x}$.  
After this 
procedure the joint quantum state displays a high degree of entanglement 
between its qubit and its wave function part. Therefore, in a final step 
we displace -- depending on the value of the qubit register -- all parts
of the wave function to the same location, so that all quantum information 
is deleted in the continuous Hilbert space and transferred to the qubit register 
(again, in an approximate sense). The controlled displacement
is done by a $\sigma_z \otimes \hat{p}$ interaction. 

Before we start the conversion process the phonon 
wave function $ \tilde{\psi}(x) $ is contained in the interval
$[- L/2, \, L/2]$. Here the length $L$ should be estimated in such a way that the
substantial part of the wave function is contained in this interval. 
We start with the following product state
% \footnote{It will become clear why 
% we would like to start with the state $ |1, \, 1, \, \dots 1 \rangle $ in the qubit register.}
\bequ
| \phi^{(0)} \rangle := |1, \, 1, \, \dots 1 \rangle \otimes | \tilde{\psi} \rangle  
\quad .
\label{starsta}
\eequ

To make subsequent procedures simpler we displace  
the wave function $ | \tilde{\psi} \rangle $ 
an amount of $L/2$ to the right such that the new wave function lies in 
the interval $[0, \, L]$. The displacement operator that achieves this is
\bequ
\exp \Big( - i \, \frac{L \; \hat{p} }{2} \, \Big) \quad.
\label{dsipop}
\eequ
As will be recalled in section \ref{secimp} we can cancel unwanted terms in eq.~(\ref{moham}) by standard
decoupling techniques by interspersing the natural evolution  with one qubit control operations.
\bequ
D_T := \exp \big(  i \, T \, c \, \sum_{j=0}^{n-1} \sigma^{(j)}_z \otimes  \hat{p} \big)
\label{defdta}
\eequ
The application of this operator to the joint 
state $ | \phi^{(0)} \rangle $ (cf.~eq.~(\ref{starsta})) of qubits 
and wave function yields
\bequ
| \phi^{(1)} \rangle := 
D_T \; \big( | 1 , \; 1 , \dots 1  \rangle \otimes  | \tilde{\psi } \rangle \big) =
| 1 , \; 1 , \dots 1  \rangle \otimes \exp \big(-i \, n \, T \, c \; \hat{p} \, \big) \; 
| \tilde{\psi } \rangle \quad .
\label{ephi1}
\eequ
A comparison with formula (\ref{dsipop}) shows that by choosing the time span 
$ T = L / (2 \; n \; c ) $ we can realize the desired displacement of the  wave 
function. The displaced state is denoted by $ | \psi  \rangle  $.

In the first part of the phase estimation algorithm the $n$ qubits are in a uniform
superposition of computational basis states and control the application 
of the operator $ \exp (2 \pi  i \; \hat{x} / L )$ to the wave function. This can 
be formulated as
\bequ
U \; \Big(\, \,  \frac{1}{2^{n/2}} \; \sum_{k = 0}^{2^n -1} \,  | k \rangle \; 
\otimes \; | \psi \rangle \; \Big) \; \; ,
\label{sche1}
\eequ
where operator $U$ is given as
\bequ
U := \exp \Big( \pha \; \sum_{j=0}^{n-1} 2^j \; P_j \otimes \frac{\hat{x}}{L} \Big) \quad .
\label{defu}
\eequ
Here $ P_j  $ is the  projection operator that acts on the $j$th qubit of the qubit register defined by 
\bequ
P_j = \openone \otimes \openone \dots \otimes |1\rangle \langle 1| \otimes \dots  \otimes \openone 
\quad \,.
\label{defproj}
\eequ
Unfortunately, we cannot implement the operator $U$ as it is with our available resources.
But since  $ P $ in eq.~(\ref{defproj}) can be written as 
$ P := ( \openone - \sigma_z ) / 2$, we can split $U$ into two factors
\bequ
U = \exp \Big( -\pi  i 
\sum_{j=0}^{n-1} 2^j \, \sigma_z^{(j)} \, \otimes \frac{\hat{x}}{L} \, \Big)
\; \; \exp \Big( \pi \, i \, (2^n-1) \, \frac{\hat{x}}{L} \, \Big)  =: \tilde{U} \; \; R \quad .
\label{factu}
\eequ
The second factor $R$ acts only on the wave function multiplying it with a position--dependent
phase. Therefore eq.~(\ref{sche1}) is equal to 
\bequ
\tilde{U} \; \Big(\, \,  \frac{1}{2^{n/2}} \; \sum_{k = 0}^{2^n -1} \,  | k \rangle \; 
\otimes \; R \; | \psi \rangle \; \Big) \; \; ,
\label{sche2}
\eequ
We can realize this transformation in the following three steps
\begin{description}
\item[i)] We first multiply the wave function $ \psi (x) $ with the phase
\bequ
\exp \Big( \pi \, i \, (2^n-1) \, \frac{x}{L} \, \Big) \; , \quad \; x \in [0, \; L] \quad \,,
\label{phasex}
\eequ
which is done as follows.
In section \ref{secimp} we will demonstrate how to realize the operator 
\bequ
\tilde{D}_T = \exp \big(-  i \, T \, c \, \sum_{j=0}^{n-1} \sigma^{(j)}_z \otimes \hat{x} \big)
\label{resbang2}
\eequ
with our available resources. Acting with $ \tilde{D}_T $ on the joint quantum state 
$ | \phi^{(1)} \rangle $ (cf.~eq.~(\ref{ephi1})) achieves the phase multiplication
\bequ
\tilde{D}_T \; \big( | 1 , \; 1 , \dots 1  \rangle \otimes  | \psi \rangle \big) =
| 1 , \; 1 , \dots 1  \rangle \otimes \exp \big(i \, n \, T \, c \; \hat{x} \, \big) \; 
| \psi  \rangle \quad .
\eequ
\item[ii)] To bring the qubit register into the uniform superposition of all computational
basis states $ | k \rangle $ we apply to each qubit the operator
\[
E := \frac{1}{\sqrt{2}} \; \left( \begin{array}{*{3} c} \;  1 \; & \; 1 \; \\ 
\; -1 \; & \; 1 \; \end{array} \right) \; ,
\] 
since
\bequ
\bigotimes_{j=0}^{n-1} E \; \, \, | 1 , \; 1 , \dots 1  \rangle  = 
\frac{1}{2^{n/2}} \; \sum_{k = 0}^{2^n -1} \,  | k \rangle \quad .
\eequ
\item[iii)] The structure of the operator $ \tilde{U} $ as defined in eq.~(\ref{factu}) is 
very similar to the one of the operator $ \tilde{D}_T $ of~eq.~(\ref{resbang2}). The only
difference are the factors $2^j$ in the exponent of $ \tilde{U} $. Clearly we cannot increase the
strength of the interaction by any selective decoupling scheme. In order to obtain
a unitary which would correspond to the exponentially growing interaction we need exponential interaction time 
(see section \ref{secimp}). 
\end{description}

After the application of these steps the quantum state of the joint system is changed to
\bequ
| \phi^{(2)} \rangle = U \; | \phi^{(1)} \rangle =
\sum_{k =0}^{2^n - 1} \, \int_0^L dx \, \Big\{ | k \rangle \otimes \frac{1}{2^{n/2}} \; 
\exp \Big(  \frac{2 \pi i \; k \; x}{L} \Big) \;  \psi (x)  \; | x \rangle \; \Big\} \quad .
\eequ
Note that we have preferred to use the notation $|x\rangle$ even though position eigenstates
do not exist (readers who appreciate mathematical rigor may forgive us).
The whole expression is nevertheless a well-defined state in the joint Hilbert space.
The second part of the phase estimation algorithm consists of the application of
an inverse Fourier transform $ {\cal F}^{-1} $ to the qubit 
register. The Fourier transformation (and its inverse) can be efficiently implemented
on a quantum computer \cite{Cop,NC}. After this transformation the quantum state 
becomes
\bea
\hspace{-.4cm} | \phi^{(3)} \rangle & = & \left( {\cal F}^{-1} \otimes \openone \right) \;
| \phi^{(2)} \rangle  \nonumber \\ 
\hspace{-.4cm} & = & \sum_{l =0}^{2^n - 1} \, \int_0^L dx \, \Big\{ \, \psi (x)  
\, | l \rangle \, \otimes \,
\frac{1}{2^{n}} \, \sum_{k = 0}^{2^n -1} \, \exp \Big[ 2 \pi  i \, k \Big( \frac{x}{L} - 
\frac{l}{2^n} \Big) \Big] \; | x \rangle \Big\} 
\label{sphi3}
\eea
$ | \phi^{(3)} \rangle $ displays a high degree of entanglement between its qubit and its
wave function part. The following procedure removes a large part 
of this entanglement and thus completes the transfer of quantum information.
For this purpose we apply the following operator to the quantum state
\bequ
V = \exp \Big(- \frac{i \, L}{2^{n+1}} \, \sum_{j=0}^{n-1} 2^j \; \sigma_z^{(j)} \otimes \hat{p} \Big) 
\quad .
\label{defv}
\eequ
As before we will discuss the implementation of this operator with our resources in 
section \ref{secimp}. Acting with operator $ V $ 
on the quantum state $ | \phi^{(3)} \rangle $ of eq.~(\ref{sphi3}),
the wave function part is displaced where the amount 
depends on the entangled qubit state $ | l \rangle $
\bea
| \phi^{(4)} \rangle & := & V \, | \phi^{(3)} \rangle :=  \sum_{l =0}^{2^n - 1} \, \int_0^L dx \, 
\Big\{  \, \psi (x)  
\, | l \rangle  \otimes  \nonumber \\
& & \frac{1}{2^{n}}  \sum_{k = 0}^{2^n -1} \, \exp \Big[ 2 \pi  i \, k \Big( \frac{x}{L} - 
\frac{l}{2^n} \Big) \Big] \, \,  | x + h  - ( l \; L ) / 2^n \; \rangle \Big\} \; ,
\eea
where we have used the quantity
\bequ
h := L \; \frac{2^n - 1}{2^{n+1}} \quad .
\label{defmidd}
\eequ
Making the substitution $ x^\prime := x - ( l \; L ) / 2^n $ we can rewrite this quantum state as
\bequ
| \phi^{(4)} \rangle = \sum_{l=0}^{2^n - 1} \, \int_{-L \; l/ 2^n}^{L \; ( 1 - l / 2^n)}  dx^\prime \, 
\Big\{ \, \frac{1}{2^{n/2}} \, \psi \Big( \frac{ l \; L }{2^n} + x^\prime \Big)  
\, | l \rangle \, \otimes \, g \Big( \frac{x^\prime}{L} \Big) \;  | h + x^\prime \rangle \; \Big\} \; ,
\label{resu1}
\eequ
where 
\bequ
g(y) := \sum_{k = 0}^{2^n -1} \, \frac{1}{2^{n/2}} \, \exp 
\left( 2 \pi  i \, k \, y \right) = \frac{1}{2^{n/2}} \; 
\frac{1 - \exp \leb \pha \, 2^n \, y \rib }{1 - \exp \leb \pha \, y \rib } \quad .
\label{defigp}
\eequ
The function $ g (y) $ is periodic with period $ 1 $. One can easily show that 
\bequ
\int_0^1 dy \, | g(y) |^2 = 1 \quad .
\label{integ}
\eequ  
Besides, the function $|g (y) |^2 $ becomes highly peaked around $ y = 0 $ for a large 
number of qubits as shown in fig.~\ref{plotg}. In fact, the width of $ | g(y) |^2 $ is proportional 
to $2^{-n}$. Outside the peak the function $ g(y) $ takes on values that have only a small modulus 
($ \propto 2^{-n/2}$). Thus the wave function part in the quantum state $ | \phi^{(4)} \rangle $
of eq.~(\ref{resu1}) displays a peak around $x^\prime = h $ where $ h $ (cf.~its definition in 
eq.~(\ref{defmidd})) is approximately the mid point of the interval $[0, \, L]$. 
\bef
\begin{center}
\mbox{\epsfig{figure=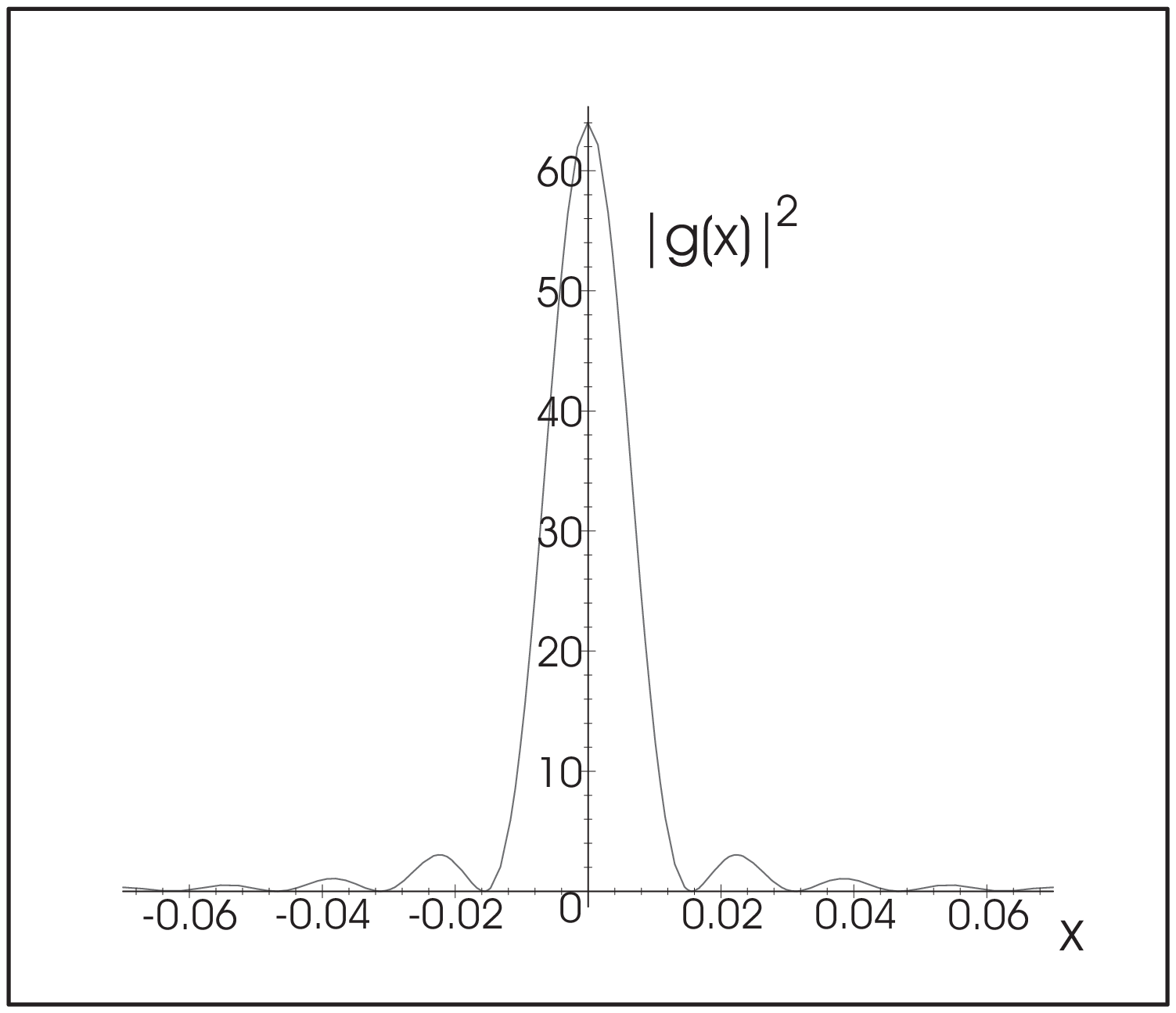, width=80mm}}
\end{center}
\caption{The function $|g(x)|^2$ for $n=6$.}
\label{plotg}
\eef

The result in eq.~(\ref{resu1}) is almost satisfactory, 
but a minor technical point should be mentioned. 
We would like that in eq.~(\ref{resu1}) each qubit register state $ | l \rangle $ has the same
"wave function" in its corresponding continuous part of the tensor space. We have
seen that the function $ g(x^\prime/L) $ can be neglected outside its peaks. Considering
the $l$--dependent integration bounds in eq.~(\ref{resu1}), the relevant wave function  
is not the full peak of $ g(x^\prime/L) $ only for low and high values of $l \in \{0, \, 
1,\, \dots , \, 2^n - 1 \} $ . Hence we have to make the additional assumption 
for the original wave function that $ \psi (x) \approx 0 $ holds in 
the two subintervals of length $ {\cal O} \leb 2^{-n} \rib $ that join the end points 
$0$ and $1$ of the interval $ [0, \, L]$.
 
Using this technical assumption and the fact that $g (x^\prime / L) $ can be 
neglected far away from its peak, we can rewrite the quantum state of eq.~(\ref{resu1}) 
to a very good approximation as   
\bequ
| \phi^{(4)} \rangle \approx
\sum_{l=0}^{2^n - 1} \, \int_{-W}^{W}  dx^\prime \, \Big\{ \, 
\frac{1}{2^{n/2}} \, \psi \Big( \frac{l \; L}{2^n} + x^\prime \Big)  
\, | l \rangle \; \otimes \; g \Big( \frac{x^\prime}{L} \Big) \;  | h + x^\prime \rangle \; \Big\} \; ,
\label{resu2}
\eequ
where $W > 0 $ is some multiple of the width 
of the function $ g(x^\prime/L) $ and hence $ W \propto 2^{-n}$.
Eq.~(\ref{resu2}) shows that now the quantum information has been transferred
to the qubit register, since the continuous Hilbert space is left with the
"standard wave function" $  g (x^\prime / L )$.
This is in accordance with the no cloning theorem \cite{NoCloning}
that precludes the copying of quantum information. 

The result eq.~(\ref{resu2}) still shows some degree of entanglement as
qubit and wave function part are connected via the integration over $x^\prime$. In order
to assess the magnitude of this entanglement, we calculate the reduced density
 operator of the qubit register
\bea
\hat{\rho} & := &
\int_{- \infty}^{ \infty} dx \; \langle x |  \phi^{(4)} \rangle \; \langle  \phi^{(4)} | x  \rangle  
\nonumber \\
& = & \frac{1}{2^n} \, \sum_{j, \, k =0}^{2^n - 1} \, \int_{-W}^{W} dx \; \Big\{ \Big| g \Big( \frac{x}{L} 
\Big) \Big|^2 \, \psi \Big( \frac{j \; L}{2^n} + x \Big) \; \bar{\psi} \Big( \frac{k \; L}{2^n} + x \Big) \Big\} \; 
| j \rangle \, \langle k | \quad .
\label{dens1}
\eea
In the following we show that for a large number of qubits this density operator can 
be replaced by the density operator
\bequ
\hat{\rho}^0 =  \frac{d^2}{2^n} \; \sum_{j, \, k = 0}^{2^n - 1} \psi \Big( \frac{j \; L}{2^n} \Big) \; 
\bar{\psi} \Big( \frac{k \; L}{2^n} \Big) \; | j \rangle \, \langle k | \quad ,
\label{rhonu}
\eequ
which is the density operator of a pure state $ | \tilde{\Psi} \rangle $ with 
\bequ  
\langle j | \tilde{\Psi} \rangle  := \frac{d}{2^{n/2}} \; \psi \leb \frac{j \; L}{2^n} \rib \; , \quad 
\quad \; j=0, \, 1, \dots , \, 2^n-1 \quad .
\label{pure1}
\eequ
The constant $ d = 1 + {\cal O} (2^{-n} ) $ ensures the correct normalization in 
eqs.~(\ref{rhonu}) and (\ref{pure1}). To demonstrate the 
possibility of replacing $ \hat{\rho} $ by $ \hat{\rho}^0 $, it is appropriate to
consider the {\em trace norm} \footnote{The trace norm of an operator $ \hat{O} $
is defined as $ \| \hat{O} \|_1 := {\rm Tr} [ ( \hat{O}^{\dagger} \, \hat{O} )^{1/2} ] $.} 
of the difference operator $ \Delta \hat{\rho} := 
\hat{\rho} - \hat{\rho}^0 $. For -- thanks to H\"older's inequality -- we can 
bound expectation values for an observable $ \hat{A} $ as
\bequ
\left| {\rm Tr} [ \, ( \hat{\rho} - \hat{\rho}^0 ) \, \hat{A} ]\right| 
\leq \| \hat{\rho} - \hat{\rho}^0 \|_1 \; \| \hat{A} \|_{\infty} \quad ,
\label{qhoeld}
\eequ
where $\| \; \; \|_{\infty} $ is the operator or spectral norm of $ \hat{A} $.
In the appendix to this article we show the following bound for the trace norm
\bequ
\| \hat{\rho} - \hat{\rho}^0 \|_1 \leq b \; \frac{1}{2^n} \; , \quad \quad n \gg 1 \; ,
\label{showbo}
\eequ
where 
\bequ
b \propto \Big( \, \int_0^1 dx \; | \psi^\prime(x) |^2 \; \Big)^{\frac{1}{2}} \quad . 
\label{l2deriv}
\eequ 
Eqs.~(\ref{qhoeld}) and (\ref{showbo}) show that in the limit of a large 
number of qubits the density operators $ \hat{\rho} $ 
and $ \hat{\rho}^0 $ are equivalent for all observables $  \hat{A} $ whose $ \| \; \; \|_2 $ 
norm diverges slower than $2^n$. We thus need a large number of qubits $n$ in order to represent
the quantum information of a wave function faithfully in a qubit register. Furthermore, according
to eq.~(\ref{l2deriv}) the accuracy of this representaion is also determined by the $L^2$ norm 
of the derivative of the wave function. Hence the smaller the derivative of the wave function,
the better works its conversion into digital information (for a fixed number of qubits). 

We have already mentioned that the time needed for the
execution of the phase estimation algorithm grows exponentially with the number of qubits.
Thus there is a trade--off between accuracy and speed for our A/D conversion algorithm. 
Squeezing operations \cite{squeez1,loudon}, i.~e.~operators of the form
\bequ
S(r) := \exp \Big( \, \frac{r}{2} \; [ a^2 - (a^\dagger)^2 ]  \; \Big) \; , \quad \; r \in \R \; ,
\label{squeeop}
\eequ
could speed up the phase estimation algorithm considerably. For, the squeezing operator could magnify 
the  wave function $ \psi (x) $ by a factor $\lambda > 1 $ while preserving its shape. This
would decrease the time for phase estimation by a factor $ \lambda^{-1} $.
Note that we can rewrite $S(r)$ as
\bequ
S(r) =\exp \Big[ i r(\hat{x}\hat{p}+\hat{p}\hat{x})  \Big]\,.
\eequ
In order to generate such unitaries we have to simulate the Hamiltonian
\[
\hat{x}\hat{p}+\hat{p}\hat{x}\,.
\]
To achieve this, we observe that
\[
[\sigma_x \otimes \hat{x}, \sigma_y \otimes \hat{p}]=
\frac{1}{2} \sigma_z \otimes (\hat{x}\hat{p}+\hat{p}\hat{x}) \quad .
\]
We conclude that $(\hat{x}\hat{p}+\hat{p}\hat{x})/2$ can 
be obtained by the following second-order simulation scheme which applies
the following $4$ Hamiltonians for a small time $\Delta T$: 
\bequ
{\rm (1)} \; \sigma_x \otimes \hat{x}  \, , \; \quad {\rm (2)} \; \sigma_y \otimes \hat{p}  \, , \; \quad
{\rm (3)} \; -\sigma_x \otimes \hat{x}  \, , \; \quad {\rm (4)} \;  -\sigma_y \otimes \hat{p} \quad .
\eequ

Up to terms $O((\Delta T)^3)$  we obtain a time evolution 
according to the desired Hamiltonian multiplied with a slow-down factor $(\Delta T)^2$ provided that
the qubit is set to the state $|1\rangle$.  
Due to
\[
({\bf 1}\otimes S(r)^k)  (\sigma_z \otimes \hat{x} ) ({\bf 1}\otimes S(-r)^k) = \sigma_z \otimes  r^k \, \hat{x}
\]
one could simulate exponentially large interaction time by a linear number of concatenated 
squeezing operations before the interaction has taken place and
undoing the squeezing afterwords. 
However, the problem with  a second-order simulation is that the running time increases with the desired accuracy.
Since the required error decreases exponentially with the desired qubits we
expect here also exponential running time. However, on a scale where squeezing operations 
are available with sufficient accuracy one could 
nevertheless expect a speed up.

\section{Selective Decoupling and Simulation of Hamiltonians} \label{secimp}
Simulation of Hamiltonians by interspersing the natural time evolution with fast control operations 
is used in NMR
since decades \cite{ernst}. These techniques are subject of many theoretical investigations 
\cite{Zanardi,VKL99,GraphPawel}.
Here we refer only to very basic ideas. 

Let $H$ be the natural Hamiltonian (\ref{jcham}).
Using the anti commutator relation between Pauli matrices
\bequ
\{ \sigma_i, \, \sigma_j \} = 0\; , \; \quad i \neq j \in \{ x, \, y, \, z \} \; ,
\label{relpaul}
\eequ
we get the equation
\bequ
\bigotimes_{j=0}^{n-1} \sigma_y \; \; \exp \big( - i \, \, \Delta T \; H \big) \;
\bigotimes_{j=0}^{n-1} \sigma_y = \exp \big( - i \, \, \Delta T \; H' \big) \; ,
\eequ
where 
\bequ
H' := c \, \sum_{j=0}^{n-1} \big( \, - \sigma^{(j)}_x   \otimes \hat{x} 
- \sigma^{(j)}_y \otimes  \hat{p} \big) \quad .
\label{effham}
\eequ
The operators $ H $ and $H'$ do not commute, but for a small time interval $ \Delta T \ll 1$ we can
use the Baker Campbell Hausdorff formula
\bequ
\exp \big( - i \, \, \Delta T \; H \big) \; \exp \big( - i \, \, \Delta T \; H' \big)
= \exp \big( - i \, \Delta T  \; ( H + H' ) \; + \, {\cal O} ( \Delta T^2 ) \, \big) \; ,
\eequ
where 
\bequ
H + H'= - 2\, c \; \sum_{j=0}^{n-1} \, \sigma^{(j)}_y \otimes  \hat{p} \quad .
\eequ 
Thus to leading order in $ \Delta T $ the unwanted 
$ \sigma^{(j)}_x \otimes \hat{x} $ terms have canceled each other.

Therefore, if during a time interval $T$ we change 
frequently between Hamiltonian evolution and the product 
of one qubit operations $ \otimes_{j=0}^{n-1} \sigma_y $, we can realize the unitary operator
\bequ
B_T = \exp \big(  i \, T \, c \, \sum_{j=0}^{n-1} \sigma^{(j)}_y \otimes  \hat{p} \big) \quad .
\label{btbt}
\eequ
In the language of \cite{GraphPawel} we have now ``simulated the Hamiltonian''
\[
c \; \sum_{j=0}^{n-1} \sigma^{(j)}_y \otimes  \hat{p}\,.
\]
In a similar way, we can also select the term with $\hat{x}$ in (\ref{jcham}) by applying $\sigma_x$-rotations to all
qubits. Complete decoupling can be achieved if we apply $\sigma_z$ to all spins since this reverses the sign
of the $\hat{x}$ and the $\hat{p}$ term. If we want to cancel all terms except from
the interaction 
\begin{equation}\label{Int}
\sigma_x^{(j)} \otimes \hat{x}
\end{equation}
for one specific qubit $j$ 
we apply $\sigma_x$ to qubit $j$ and $\sigma_z$ to all the other qubits. 
To simulate the time evolution
\[
\exp\Big(i \, T \sum_j 2^j\sigma_x^{(j)}\otimes \hat{x} \Big)
\]
we may concatenate the commuting unitaries
\[
\exp\Big(i \, T 2^j\sigma_x^{(j)}\otimes \hat{x} \Big)\,,
\]
which are obtained by applying the simulated interaction (\ref{Int}) on a time interval of length $2^j T$.
If we apply arbitrary single qubit unitaries $U_j$ to  qubit $j$  initially and apply $U_j^\dagger$ afterwords,
we obtain the time evolution 
\[
\exp\Big(i \, T 2^j (U_j \sigma_x^{(j)} U_j^\dagger ) \otimes \hat{x} \Big)\,.
\]
This shows that we can replace the  Pauli matrix $\sigma_x$ in  eq.~(\ref{Int}) 
by other Pauli matrices or by $-\sigma_x$ as we like.

\section{Digital-Analogue Conversion}

\label{secDA}

Using the results of the previous sections, we can easily describe an algorithm for digital-analogue
conversion. Roughly speaking, the argument is as follows.
Let $Z$ be the transformation on the continuous and the discrete degrees of freedom
which implements the complete analogue-digital conversion algorithm.   
Provided that the wave function $\psi$ was sufficiently smooth   
the system ends up almost in the product state 
\begin{equation}\label{Prod}
|\varphi \rangle \langle \varphi |\otimes |\tilde{\Psi}\rangle \langle \tilde{\Psi} |
\end{equation}
where 
$| \tilde{\Psi}  \rangle $ is a superposition state with coefficients $\psi (j \; L / 2^n )$ with $j=0, \, 1 , \, 
\dots,2^n-1$ as defined in eq.~(\ref{pure1}) and $\varphi$ 
is the function whose absolute square was 
plotted in fig.~\ref{plotg} translated by $L/2$.
If we apply $Z^\dagger$ to the state  (\ref{Prod})
we obtain hence almost the original wave function $\psi$.
It is clearly required that the values $\psi(j \; L /2^n)$ correspond to 
some sufficiently smooth wave function. We have explained that sufficiently smooth means here 
that $|\psi'(x)|$ is small enough. Hence the vector $|\tilde{\Psi}\rangle$  corresponds
to a smooth wave function whenever the values $\psi(j \; L / 2^n )$ do not
vary too much for adjacent $j$. 

In order to obtain definite accuracy bounds from this idea we recall that the reduced state $\hat{\rho}$ of the 
discrete register satisfies
\bequ
\| \hat{\rho} -|\tilde{\Psi}\rangle \langle \tilde{\Psi} |\, \|_1 \leq b \frac{1}{2^n}=:\epsilon\,.
\eequ
If we set $ \hat{A} = |\tilde{\Psi}\rangle \langle \tilde{\Psi}|$ 
in eq.~(\ref{qhoeld}), we obtain
\bequ
\| \hat{\rho} -|\tilde{\Psi}\rangle \langle \tilde{\Psi} | \|_1 \geq  
| \langle \tilde{\Psi} | \hat{\rho} |\tilde{\Psi}\rangle -1 | \,.
\eequ
Hence the largest eigenvalue of $\hat{\rho}$ is at least $1-\epsilon$.
Let 
\bequ
|\phi\rangle=\sum_j c_j |\alpha_j \rangle \otimes |\beta_j\rangle
\eequ
be the Schmidt decomposition of the exact bipartite state after the $A/D$ conversion. Then the absolute
square $|c_0|^2$  of the dominating coefficient $c_0$ is the largest eigenvalue of the reduced density operators
on both subsystems. Hence the square of the norm distance between 
$|\phi\rangle$ and $c_0 |\alpha_0\rangle \otimes |\beta_0\rangle$ is at most $\epsilon$.  
It follows that the error which arises from replacing the joint state by the tensor product state 
\bequ
|\alpha_0\rangle \langle \alpha_0|\otimes |\beta_0\rangle \langle \beta_0| 
\eequ
is
of the order $\epsilon$. We know furthermore that we may replace the state of the digital
system by $|\tilde{\Psi}\rangle$ such that the error  in trace norm is in the order of $\epsilon$.
We conclude that applying $U^\dagger$ to
\bequ
|\alpha_0\rangle \langle \alpha_0| \otimes |\tilde{\Psi}\rangle \langle \tilde{\Psi} | 
\eequ
leads to the wave function $\psi$ up to a trace norm error in the order of $\epsilon$. 
The initialization of the continuous system to the state $|\alpha_0\rangle$ is clearly a non-trivial task.
It is some wave packet which is similar to the function $g$ translated by $L/2$. 
To construct algorithms which work also for more general initializations shall not be our subject here.

\section{Conclusions} \label{secconc}
In this article a quantum algorithm has been outlined that can read in the quantum
information of a wave function into a qubit register. This can be viewed as
the quantum analogue of an A/D converter which encodes an $N$-dimensional system 
obtained by discretization of a continuous wave function into a $\log_2 N$ qubit register.

The principal ingredient for the A/D converter is the use of the phase estimation 
algorithm that is already popular for other purposes in quantum computing.
The principal resources are interaction Hamiltonians which are  
tensor products of a position or a momentum operator with a Pauli matrix of the discrete system.
Effective Hamiltonians of this type can for instance be obtained from the Jaynes Cummings Hamiltonian
using standard techniques for selective decoupling.

Whether the proposed A/D and D/A converters could be realized depends on the one hand on
the experimental progress. On the other hand and perhaps even more important, further 
theoretical work is required to investigate more systematically the advantages that 
arise when one combines discrete and continuous degrees of freedom in quantum computing.
\section*{Acknowledgment}
The authors are grateful to the Landesstiftung Baden--W\"urttemberg that supported
this work in the program ''Quantum-Information Highway A8'' (project ''Kontinuierliche Modelle
der Quanteninformationsverarbeitung'').
This work has been completed during DJ's visit of Hans Briegel's group
at IQOQI in Innsbruck, whose hospitality is gratefully acknowledged.

\appendix
\section*{Appendix} \label{proof}
In this appendix we prove eq.~(\ref{showbo}). To simplify the notation we set $L=1$.
We start with the density operator $ \hat{\rho} $ and use the mean value theorem 
in eq.~(\ref{dens1}) 
\bequ
\psi \Big( \frac{j}{2^n} + x \Big) = \psi \Big( \frac{j}{2^n} \Big) 
+ x \; \psi^\prime \Big( \frac{j}{2^n} + \xi(j, \, x) \Big) \; \;  {\rm with} \; \; 
0 \leq | \xi(j, \, x)  | \leq |x|\leq W 
\label{meanv}
\eequ
and analogously for the conjugate function $ \bar{\psi} ( k/2^n + x ) $.
Therewith we can write the matrix elements of the difference operator $ \Delta \hat{\rho} 
= \hat{\rho} - \hat{\rho}^0 $ in the standard basis as \footnote{In the 
limit $n \to \infty $ the normalization factor $ d^2 = 1 + {\cal O} (2^{-n} ) 
$ can be neglected.}
\bea
\leb \Delta \rho \rib_{j \, k} = \frac{1}{2^n} \; \int_{-W}^{W} dx \; 
\Big\{ | g(x)|^2 \, x \,\Big[ & & \hspace{-.8cm} \psi \Big( \frac{j}{2^n} \Big) \; 
\bar{\psi}^\prime \Big( \frac{k}{2^n} + \xi(k, \, x) \Big) \nonumber \\ 
& + & \hspace{-.2cm} \bar{\psi} \Big( \frac{k}{2^n} \Big) \; \psi^\prime \Big( \frac{j}{2^n} + \xi(j, \, x) \Big) 
\Big] \Big\}
\eea
We can rewrite this expression in an operator form 
\bequ
\Delta \hat{\rho} = \frac{1}{2^n}  \; \int_{-W}^{W} dx \; \Big\{ | g(x)|^2 \, x \,
\Big[ | \, \Psi \rangle \, \langle \Psi^\prime_x \, | + 
| \, \Psi^\prime_x \rangle \, \langle \Psi \, | \, \Big] \, \Big\} \; ,
\label{opfor}
\eequ
where the $2^n$ dimensional vectors $  | \, \Psi \rangle $ and $  | \,  \Psi^\prime_x \rangle $ are
defined in the standard basis as
\bequ
\langle j \, | \,  \Psi \rangle := \psi \Big( \frac{j}{2^n} \Big) \; , \quad \; 
\langle j \, | \, \Psi^\prime_x \rangle := \psi' \Big( \frac{j}{2^n} + \xi(j, \, x) \Big) \quad .
\eequ
Using the triangle inequality, eq.~(\ref{opfor}) leads to the following estimate for the trace 
norm  
\bequ
\| \Delta \hat{\rho} \|_1 \leq \frac{1}{2^n} \, \Big( \, \int_{-W}^{W} dx \; 
| g(x)|^2 \, | x | \; \Big) \; \sup_{-W \leq x \leq W} \big\{ \| \; | \, \Psi \rangle \, 
\langle \Psi^\prime_x \, | + | \, \Psi^\prime_x \rangle \, \langle \Psi \, | \; \|_1 \big\} 
\quad .
\eequ
We now consider the trace norm of the operator 
\[
\hat{B}^x :=  | \, \Psi \rangle \, \langle \Psi^\prime_x \, | + 
| \, \Psi^\prime_x \rangle \, \langle \Psi \, | \quad .
\]
We see from this formula that the operator acts non--trivially only on the two dimensional 
subspace ${ \rm span}\{ \, | \, \Psi \rangle , \,  | \, \Psi^\prime_x \rangle \, \}$. Choosing an ONB
$ \{ | \, 0 \, \rangle, \,  | \, 1 \, \rangle \} $ for this subspace with $ | \, 0 \, \rangle := | \, 
\Psi \rangle / \|  | \, \Psi \rangle \, \|_2$ ( $ \| \; \; \|_2 $ is the Euclidean norm), 
we arrive at the following matrix representation for $ \hat{B}^x $ 
\bequ
B^x = \| \,  | \, \Psi \rangle \, \|_2 \; \, \| \, | \, \Psi^\prime_x \rangle \, \|_2 \; 
\left( \begin{array}{*{3} c} \; \langle e_x \, | \,  0 \, \rangle + \langle \, 0 \, | \,  
e_x \rangle \; & \; \; \langle e_x \, | \,  1 \, \rangle \; \\ \;  \langle \, 1 \, | \,  e_x 
\rangle \; & \; \; 0 \; \end{array} \right) \; ,
\label{matrb}
\eequ
where $ | \, e_x \rangle := | \, \Psi^\prime_x \rangle / 
\| \, | \, \Psi^\prime_x \rangle \, \|_2 $ is a unit vector. From eq.~(\ref{matrb})
the trace norm of $ \hat{B}^x $ can be bounded uniformly
\bequ
\| \, \hat{B}^x \, \|_1 \leq  \| \,  | \, \Psi \rangle \, \|_2 \; \, \| \, | \, 
\Psi^\prime_x \rangle \, \|_2 \; \sum_{j, \, k = 0}^1 \left| B^x_{j \, k} \right| \leq 
 4 \; \| \,  | \, \Psi \rangle \, \|_2 \; \, \| \, | \, 
\Psi^\prime_x \rangle \, \|_2 \; \; .
\label{boundb1}
\eequ
In addition, for $n \gg 1$ we can evaluate the Euclidean norms of the vectors $ | \, 
\Psi \rangle $ and $ | \, \Psi^\prime_x \rangle $ approximately as
\bequ
\|  | \, \Psi \rangle \, \|_2 \approx 2^{\frac{n}{2}} \; , \quad \;
\| \, | \, \Psi^\prime_x \rangle \, \|_2 \approx 2^{\frac{n}{2}} \; 
\Big( \, \int_0^1 dx \; | \psi^\prime(x) |^2 \; \Big)^{\frac{1}{2}} \quad .
\label{estvec}
\eequ
Using the results of eqs.~(\ref{boundb1}) and (\ref{estvec}), we can bound 
the difference between the two density operators as
\bequ
\| \Delta \hat{\rho} \|_1 \leq 4 \; \Big(  \, \int_{-W}^{W} dx \; 
| g(x)|^2 \, | x | \; \Big) \; \Big( \, \int_0^1 dx \; | \psi^\prime(x) |^2 \; 
\Big)^{\frac{1}{2}} \quad .
\label{sumbo}
\eequ
Since for $n \gg 1$ the width $ W \propto 2^{-n} $ and thus
\bequ
\int_{-W}^{W} dx \; \{ \, | g(x)|^2 \, | x | \, \} \propto 2^{-n} \; ,
\eequ
the inequality (\ref{sumbo}) establishes the bound eq.~(\ref{showbo}) with the constant
\bequ
b \propto \Big( \, \int_0^1 dx \; | \psi^\prime(x) |^2 \; \Big)^{\frac{1}{2}} \quad . 
\eequ

\end{document}